# A novel cost-effective fabrication of a flexible neural probe for brain signal recording

Alireza Irandoost[1], Amirreza Bahramani[1,2], Roya Mohajeri[1], Faezeh Shahdost-Fard[3], Ali Ghazizadeh[1], Mehdi Fardmanesh[1*]

[1] Department of Electrical Engineering, Sharif University of Technology, Tehran 1458889694, Iran.

[2] School of Cognitive Sciences, Institute for Research in Fundamental Sciences (IPM), Tehran 1458889694, Iran.

[3] Department of Chemistry Education, Farhangian University, Tehran 14665889, Iran.

[*] To whom correspondence should be addressed: Mehdi Fardmanesh, Email: fardmanesh@sharif.edu

**Abstract**

This study introduces a novel, flexible, and implantable neural probe using a cost-effective microfabrication process based on a thin polyimide film. Polyimide film, known as Kapton, serves as a flexible substrate for microelectrodes, conductive tracks, and contact pads of the probe, which are made from a thin film of gold (Au). SU-8 is used to cover the corresponding tracks for electrical isolation and to increase the stiffness of the probe for better implantation. To evaluate the performance of the fabricated probe, electrochemical impedance spectroscopy (EIS) and artificial neural signal recording have been used to characterize its properties. The microelectrode dimensions have been carefully chosen to provide low impedance characteristics, which are necessary for acquiring local field potential (LFP) signals. The in vivo LFP data have been obtained from a male zebra finch presented with auditory stimuli. By properly filtering the extracellular recordings and analyzing the data, the obtained results have been validated by comparing them with the signals acquired with a commercial neural electrode. Due to the use of Kapton, SU-8, and Au materials with non-toxic and adaptable properties in the body environment, the fabricated neural probe is considered a promising biocompatible implantable neural probe that may pave the way for the fabrication of other neural implantable devices with commercial aims.

# 1. Introduction

Understanding the neural correlates of behavior is a main goal in neuroscience. To do so, one needs to effectively record the neural activity. Two classes of information carrying electrical signals, including action potentials (spikes) and local field potentials (LFPs) are recorded by utilizing a neural interface in the brain [1]. Neural recording probes have significantly revolutionized neurophysiology research by extracellularly detecting low-frequency LFP oscillations and high-frequency action potentials of single units. Over the past few decades, various neural probes with diverse materials and structures, based on sapphire, glass, silicon and metals, have been developed for neuronal activity recordings [2-5]. However, the applicability of these rigid recording units to relatively large sizes is limited by severe tissue damage and inflammatory responses in the brain tissue. Conventional rigid metal wire electrodes and silicon-based neural probes have a higher elastic modulus than brain tissue [6]; the resulting mechanical mismatch between the stiff probe and soft brain tissue movement is not compatible with long-term implantation. Furthermore, histological investigation of intra-cortical tissue exhibits that the activated microglial and glial cells hinder local neuron regeneration at implanted sites and disrupt the recording process by forming a compact and insulating sheath around rigid probes [7]. These effects also reduce the signal-to-noise ratio which impacts the quality of the spike sorting of the recorded units.

To address these problems, some flexible probes integrated on polymer substrates such as liquid crystal polymer (LCP) [8], polyimide (PI) [9], poly-(para-chloro-xylylene) (Parylene C) [10] and polydimethylsiloxane (PDMS) [11] have provided more compliant probe-brain interfaces. Additionally, they prevented a decrease in the signal-to-noise ratio by providing a better mechanical match to the tissue and deforming in response to tissue movement. Nevertheless, the slender shanks of the excessively flexible polymer-based probes are limited for superficial neural signal recording (< 3 mm deep) because they are susceptible to mechanical buckling or rupturing during insertion into brain tissue due to some localized stress [12]. Furthermore, utilizing these expensive materials restricts the length and number of electrode sites due to the complex process of layer assembling and the necessity of an additional carrier for chip integration and an electrical interconnect. Thus, the development of a cost-effective and flexible implantable neural probe with a small-feature geometry and high spatiotemporal resolution for long-term stability improvement is still demanded for further neurophysiological experiments.

The effective parameters in building flexible platforms with small interconnection lines are affected by several restrictions. For example, various processes, such as deposition and patterning of the masking layer, baking, and wet etching using chemicals, may alter the physicochemical properties and the nature of the material as a sub-layer during the fabrication process. Thus, selecting a suitable material compatible with the applied chemical conditions is the main bottleneck in the manufacturing of a flexible neural probe. Kapton (Fig. 1 (a) and (b)), a robust and biocompatible PI-based material [13] with electrical, chemical and mechanical stability against extreme temperature, vibration and other demanding environments [14-15], may be a beneficial candidate for the fabrication of a low-cost flexible neural probe for in vivo investigations. This non-toxic material supports suitable cell adhesion, proliferation, and differentiation, and does not significantly induce oxidative stress; these advantages of Kapton are pivotal for long-term implantation aims [16-17]. SU-8 (Fig. 1 (c)), an epoxy-based photoresist, provides durable and chemically stable microstructures with high aspect ratios and results in good compatibility with conventional micromachining techniques such as photolithography and spin-coating processes. Due to its favorable mechanical properties and ability to support cell growth, resulting in good cell viability, with minimal localized adverse effects in in vivo implant studies, SU-8 is considered a biocompatible material [18-19]. These advantageous properties arise from improved Young's modulus and high aspect ratio capability for good dimensional control and patterning of narrow probes with vertical sidewalls (< 100 μm) for clean insertion into soft tissue [18], [20], [21]. Additionally, its high uniformity and adhesion properties ensure the positioning of the metallic tracks on the surface of the Kapton polymer layer for in vivo studies [22].

This study introduces an innovative and low-cost microfabrication technique for the affordable manufacturing of the implantable neural probe with four recording gold (Au) sites as the microelectrodes. The fabrication process is based on the cutter plotting of Kapton film and using the photolithography method via spin coating of SU-8 on a flexible Kapton layer. Using polyimide film instead of photosensitive polyimide resin reduces fabrication costs, steps, and the use of hazardous materials. The applicability of the developed neural probe has been investigated with both artificially simulated spikes and in vivo testing for LFP signals in anesthetized zebra finch. Furthermore, the obtained results compared with a commercial wire electrode to evaluate proficiency of the manufactured neural probe in recording LFP signals.

## 2. Experimental section

### 2.1. Materials, instruments, and software

Kapton film with a thickness of 50 µm (DuPont), SU-8 photoresist (Kayaku, 2005), 3.5% sodium hydroxide solution (Merck), as well as hydrochloric acid and nitric acid (Merck) were used in the fabrication process. Isoflurane and dental cement were used during surgery.

A cutter plotter device (Graphtec, CE7000) was employed to cut the Kapton layer to the desired size. The Au deposition was performed using a physical vapor deposition system (PVD, Denton Vacuum Desk IV). Patterning of the microelectrodes for the photolithography process and imaging of the microelectrodes at various magnifications were carried out using a mask aligner (Karl Suss MJB3 UV200). Simulated neural spikes were generated using a Blackrock digital neural signal simulator (Blackrock), and signal recording was done with a Blackrock signal recorder. Morphological imaging of the Au layer on the Kapton substrate was conducted with a field emission scanning electron microscope (FESEM, TESCAN, MIRA 3). Electrochemical impedance measurements were carried out using an Autolab electrochemical workstation (Metrohm, PGSTAT302N).

CorelDRAW 2024 software was utilized to design the photolithography mask, and MATLAB 2024b was employed for signal processing and figure plotting.

### 2.2. Design and fabrication

Dimension design of the flexible neural probe is a critical step for low-damage implantation into deep brain tissue. According to our strategy, this probe is designed with a single microneedle with a length, width, and thickness of 5 mm, 250 µm, and 100 µm, respectively, as depicted in Fig. 1 (d). The needle is embedded with four microelectrodes with a 30 µm diameter, and the distance between the two microelectrodes is 100 µm. The needle widens at the end, where there are four large contact pads for each microelectrode to be wired to the readout circuit.

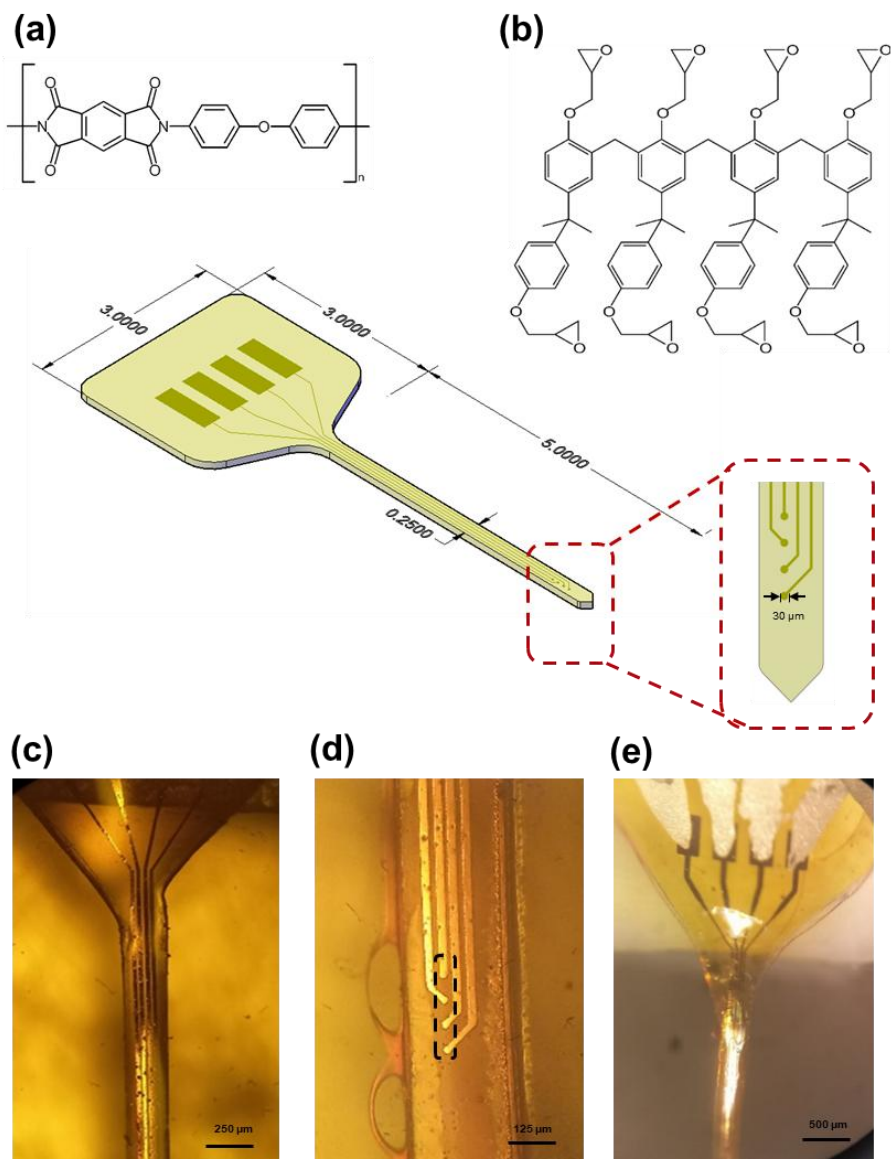

**Fig. 1. (a)** Chemical structure of Kapton **(b)** Chemical structure of SU-8 photoresist. **(c)** Isometric view of the designed neural probe (the unit of annotated numbers is in millimeters) **(d)** optical microscopy image of the probe with embedded four traces connecting microelectrodes to contact pads, **(e)** four microelectrodes and **(f)** contact pads and tracks after mounting.

The Kapton film with a thickness of 50 μm was cut with a cutter plotter device to define the geometry of the designed neural probe. The resulting flexible substrate was attached to a glass and then a 200 nm of Au was deposited on it using physical vapor deposition. Thanks to the good adhesion of Au on Kapton film, a further intermediate adhesion layer, such as titaniu006D, was not needed. Afterward, a 1 μm thick layer of positive photoresist (Shipley S1813) was spun on the Au-deposited substrate and the electrode pattern was created on the probe with a standard photolithography system based on the designed mask. Subsequently, the SU-8 layer was spun-coated on the processed surface with a thickness of 5 µm as an insulating layer. The Au tracks were covered with an insulating layer except for the electrode sites and the contact pads. Finally, to increase the stiffness of the fabricated probe, the backside of the probe was spun-coated with 20 µm of SU-8. For packaging and wiring, the neural probe was attached to a holder for connecting the contact pads to the external readout circuit. The fabrication process of the probe is schematically shown in the scheme. 1.

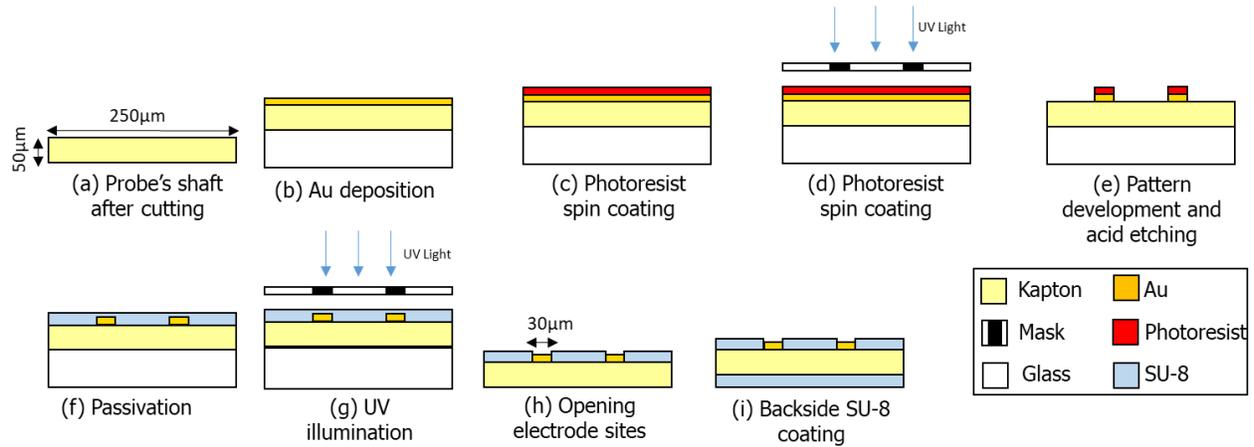

**Scheme. 1. The schematically fabrication process of implantable neural probe (figures are not scaled for clear illustration)**: **(a)** cross section of Kapton after cutting, **(b)** Au deposition, **(c)** photolithography, **(e)** pattern development and acid etching, **(f)** passivation, **(g)** UV illumination, **(h)** opening electrode sites and removing glass **(i)** backside SU- 8 coating.

## 3. Results and discussion

### 3.1. Structural and morphological features of the neural probe

The proposed neural probe is manufactured using a biocompatible and low-cost Kapton film as the flexible initial layer. This durable platform not only embeds a chemically and physically stable surface without toxicity for further processing but also decreases fabrication costs. As evidenced in Fig. 2 (a) and (b), the images of the optical microscopy for the fabricated neural probe reveal the well-structured characterization of the designed needle with embedded four traces connecting microelectrodes to contact pads, along with four microelectrodes. The FESEM image of the 200 nm Au on Kapton in Fig. 2 (d) shows the roughness of the surface. As shown in Fig. 2 (e) increasing the thickness of Au reduces the roughness of the surface.

## 3.2. Electrochemical investigation of the neural probe

The EIS technique was applied to study the electrochemical behavior of the fabricated microelectrodes. To do so, a three-electrode system consisting of the microelectrodes of a neural probe as the working electrode, Ag/AgCl (3 M KCl) as the reference electrode, and the platinum wire as the counter electrode was used (Fig. 3 (a)). The impedance of the electrode was recorded at a root-mean-square (RMS) sinusoidal potential of 10 mV with frequencies ranging from 100 Hz to 100 kHz in a physiological saline solution. As indicated in Fig. 3 (b), the impedance-frequency curve reveals that the signal is decreased with increasing frequency, which can be attributed to the effect of the double-layer capacitance ($C_{dl}$). The impedance value of the designed microelectrodes at 1 kHz is obtained to be 440 kΩ. Thanks to the relatively high roughness of the Kapton surface, as shown in Fig. 2 (d), the deposited Au layer takes the shape of the substrate. These feature decreases the impedance of the electrode because the effective surface area of the Au layer has been increased.

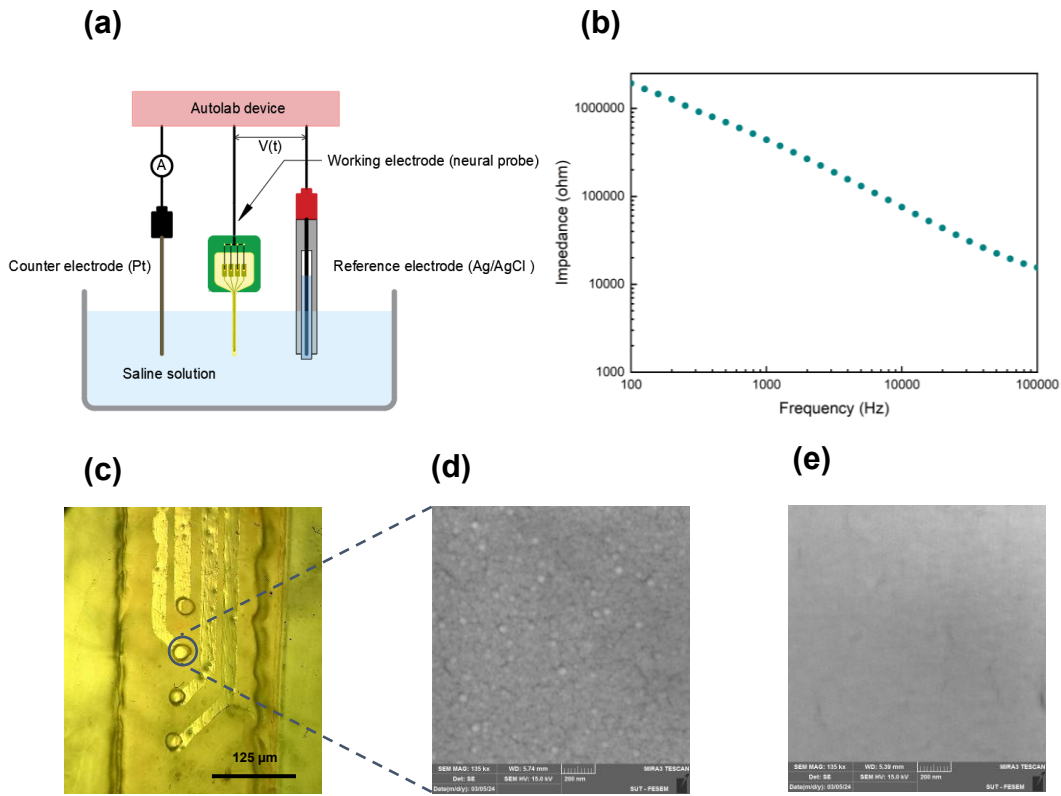

**Fig. 2. EIS measurement and morphological features of the neural probe (a)** A Schematic representation of the electrochemical set-up based on a three-electrode system for EIS studies and **(b)** the obtained impedance-frequency curve in the electrochemical system at an RMS sinusoidal potential of 10 mV with frequencies ranging from 100 Hz to 100 kHz in a physiological saline solution. **(c)** optical image of electrodes **(d)** FESEM image of the 200 nm and **(e)** 400 nm Au deposited on Kapton film in the fabricated neural probe.

## 3.3. Recording the simulated neural signal

To evaluate the applicability of the fabricated neural probe in signaling, it was tested with a neural signal simulator and the quality of recorded signals was examined. Therefore, Blackrock digital neural signal simulator created some simulated spikes and the signal was transmitted through a stainless-steel wire inserted deeply into a physiological saline solution (0.9% sodium chloride). The saline solution has an electrical conductivity of 12 mS cm$^{-1}$ at 20 ºC, which is due to the mobile ions allowing electron charge transfer through the solution [20]. This conductivity provided a low impedance medium for transferring signals. The distance between the stimulating tip and the recording probe was long enough to ensure that the signal could be received. The recorded signals using the fabricated probe were filtered with a band-pass filter between 300 Hz to 7 kHz to detect all the spikes of the generated signal by the Blackrock simulator system. Fig. 3 (a) shows the raw data captured by the fabricated neural probe and Fig. 3 (b) depicts the power spectrum of it, showing peak power at 2.4 kHz due to the presence of a neural signal. The filtered signals and the related single spike are demonstrated in Fig. 3 (c) and (d), respectively. The individual impulses of the signal in Fig. 3 (d) resemble the spikes from real neurons. According to Fig. 3 (d), the waveform has a biphasic shape to simulate the principal ion exchange across the membrane of a cell.

## 3.4. Neural signal recording and analysis

The surgical procedures were conducted on a male zebra finch (*Taeniopygia guttata)* and all procedures were approved by the Research Ethics Committee of the School of Cognitive Sciences (SCS) of the Institute for Research in Fundamental Sciences (IPM, protocol number 1402/40/1/2841). Initially, the finch was administered anesthesia through an isoflurane/oxygen flow. The concentration of isoflurane was gradually increased from 0.4% per minute until it reached the maximum level of 2%. Afterward, the anesthetized bird was positioned onto a stereotaxic device, where a steady flow of isoflurane/oxygen was delivered through a face mask fixed at 1.4%. Subsequently, the feathers on the head were plucked and the skin was incised. Following that, a craniotomy was conducted using a dental drill and an aperture for the insertion of a stainless-steel screw to establish a ground connection. Next, a craniotomy was carried out above the HVC (used as a proper name for vocal motor nucleus in the nidopallium), a brain region that plays a crucial role in the learning and production of songs in zebra finches (Fig. 4 (a)) [23].

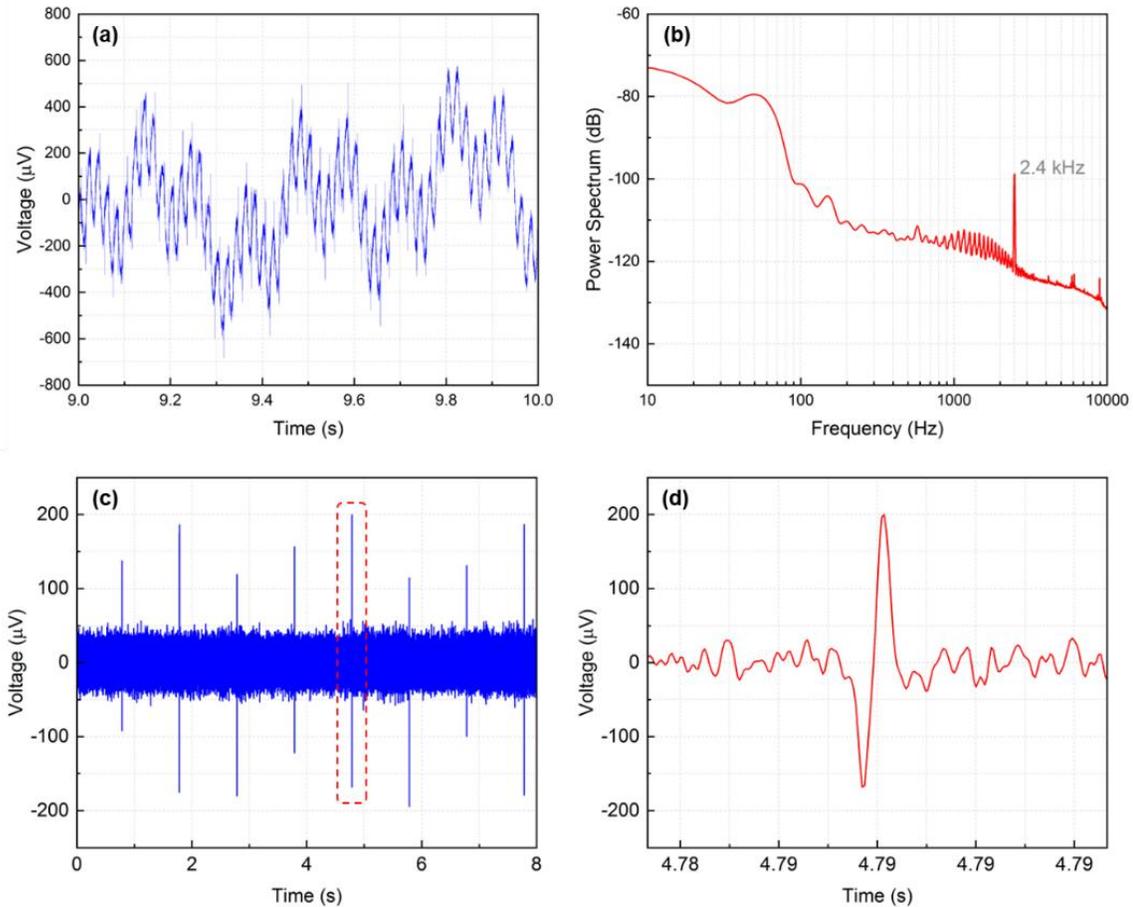

**Fig. 3. Recording simualted neural signal (a)** Raw signal and **(b)** its power spectrum, **(c)** filtered signal and **(d)** the corresponding single spike for the fabricated probe during the simulation process by a neuronal signal simulator after applying a band-pass filtering between 300 Hz to 7 kHz in 0.9% W/V of sodium chloride.

Extensive previous research has shown that, in adult male birds under anesthesia, HVC projecting neurons respond to certain auditory stimuli, including birdsong [24-25]. Once the HVC area was identified using stereotaxic coordinates (2.0 mm lateral and 0.1 mm anterior), the dura mater was removed, and a manipulator was used to insert the probe into the brain tissue, as shown in Fig. 4 (b). The Blackrock neural recording system was used for acute recordings with a 30 kHz sampling rate. First, the probe was lowered in search of LFP patterns, and once LFP patterns were detected in the HVC region, stimuli were presented. Four distinct types of vocal stimuli, including white noise, pure tone, and two different types of conspecific zebra finch songs, were used and each was played 22 times for the subject. Here, one of the conspecific bird songs was used to analyze the recorded LFP signal and compare it to the previously recorded LFP from a commercial tungsten electrode. The presented song waveform and its corresponding spectrogram and the mean LFP response recorded with tungsten wire electrode are displayed in Fig. 4 (c), (d), and (e) respectively.

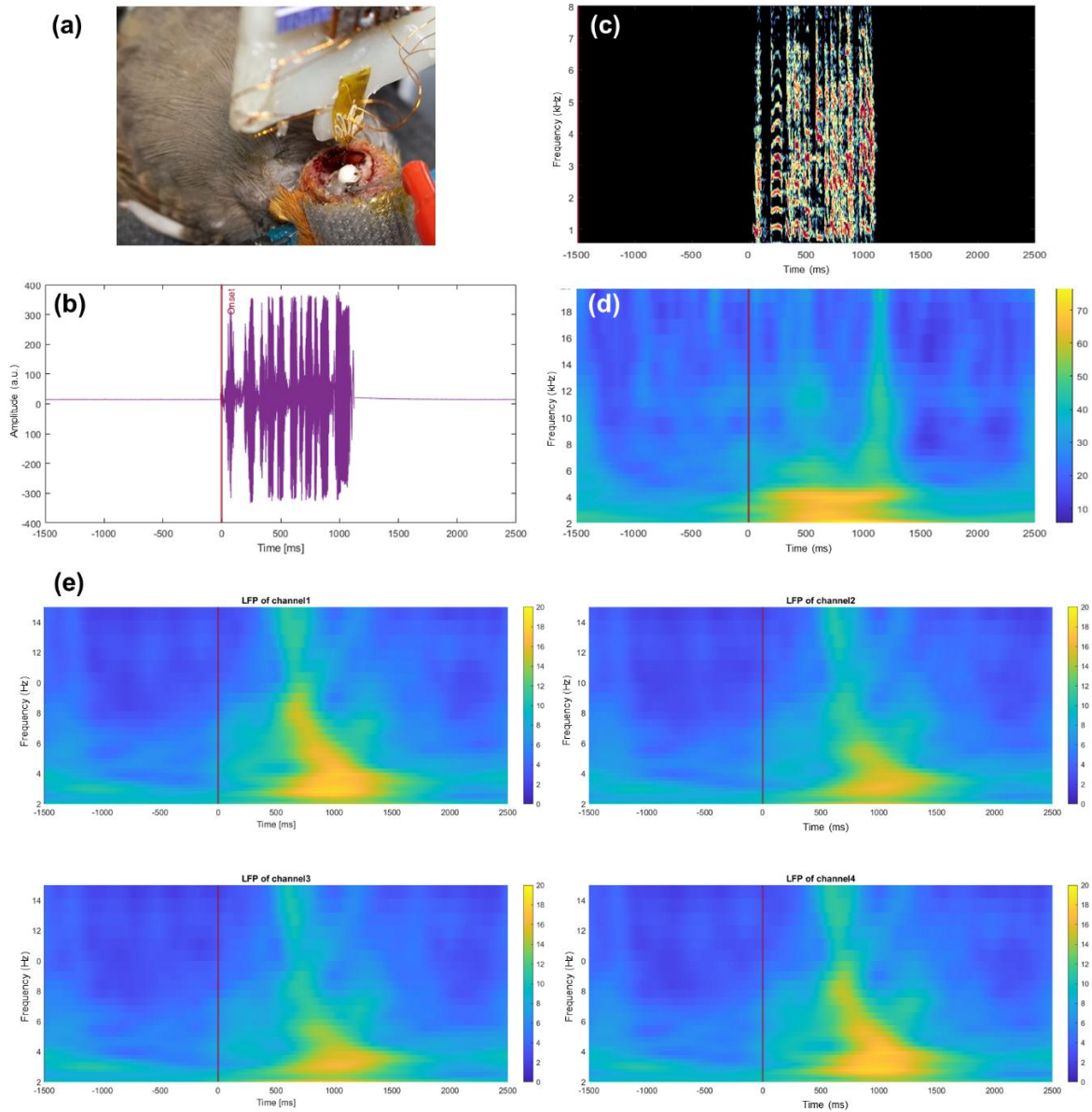

**Fig. 4. The response of the LFP signal to the presented song stimulus (a)** The implantation image of the fabricated neural probe into the brain of the Zebra finch. **(b)** The bird song waveform played during recording the brain signal and **(c)** its corresponding spectrogram. **(d)** The mean LFP response across 22 trials obtained from the commercial tungsten electrode. **(e)** The mean LFP response across 22 trials, obtained from the four channels of the fabricated probe

To analyze the LFP signal, the raw data was filtered using a low-pass filter with a cutoff frequency of 300 Hz. Then, the signal was transferred to the frequency domain using a Morse wavelet with symmetry parameter gamma (γ) equal to 3 and time-bandwidth product equal to 60. Fig. 5 (f) presents the LFP response to the presented song stimulus, which was averaged across 22 trials for 4 recording channels. The figure clearly shows sustained activity in the Delta (0.5-4 Hz) and Theta (4-7 Hz) frequency bands during the presentation of the song. This observation is consistent with previous reports that demonstrate the response of HVC neurons to conspecific songs [26]. Notably, the duration of response was similar to that of the presented song, approximately 1200 ms. To ensure the validity of the results, the experiment was repeated using a commercial tungsten electrode. Fig. 4 (e) shows a nearly identical LFP response pattern, confirming the reliability of the data obtained with the fabricated neural probe. The difference in LFP power values between them is due to the different probe impedances, positions, and two distinct birds. Table .1 compares the neural probe with tungsten micro wire electrode. The fabricated microelectrodes are suitable for LFP recordings due to their slightly larger surface area. However, this feature makes it difficult to detect spikes during in vivo experiments.

**Table 1.** Comparison of commercial metal wire electrode and Kapton neural probe

| Type of the microelectrode | Substrate | Electrode material | Impedance at 1 kHz | Insulator | Number of electrodes |
|---|---|---|---|---|---|
| Metal wire electrode (commercial electrode) | Rigid | Tungsten | ≈ 900 KΩ | Glass | 1 |
| Kapton neural probe (This work) | Flexible | Au | 440 KΩ | SU-8 | 4 |

## 4. Summary and conclusion

In this work, a flexible neural probe for extracellular neural recording was made using a low-cost microfabrication technique. Kapton film, also known as polyimide film, was used as the sub-layer in the fabrication process, and SU-8 was used to provide an insulating layer and to increase the probe's stiffness. Due to the utilization of non-toxic and adaptable materials with the body environment, such as Kapton, SU-8 and Au materials, the developed neural probe is considered a biocompatible and cost-effective implantable platform for deep-brain signal recording for commercial purposes. To evaluate the performance of the manufactured neural probe in a real-world context, we performed surgery on a male zebra finch and presented vocal stimuli to it. The LFP of the recorded neural activity was analyzed, which demonstrated a consistent response to the presented song stimulus. We also validate our results by comparing them to the results obtained from a commercially available tungsten electrode. The comparison showed a high degree of similarity between the two LFP responses, indicating that the developed microelectrodes are capable of properly recording neural activity.